\documentclass[aps,prb,twocolumn,floatfix,groupedaddress]{revtex4}

\usepackage{graphicx}
\usepackage{dcolumn}
\usepackage{bm}

\begin{document}

\title{Spin-dynamics simulations of the triangular 
antiferromagnetic $XY$ model}   

\author{Kwangsik Nho and D. P. Landau}

\affiliation{Center for Simulational Physics, University of Georgia, Athens, Georgia 30602}

\date{\today}

\begin{abstract}
Using Monte Carlo and spin-dynamics methods, we have investigated the dynamic 
behavior of the classical, antiferromagnetic $XY$ model on a triangular 
lattice with linear sizes $L \leq 300$. The temporal evolutions of 
spin configurations were obtained by 
solving numerically the coupled equations of motion for each spin using 
fourth-order Suzuki-Trotter decompositions of exponential operators. 
From space- and time-displaced spin-spin correlation functions and their
space-time Fourier transforms we obtained the dynamic structure factor
$S({\bf q},w)$ for momentum ${\bf q}$ and frequency $\omega$. 
Below $T_{KT}$(Kosterlitz-Thouless transition), both the in-plane 
($S^{xx}$) and the out-of-plane ($S^{zz}$) components 
of $S({\bf q},\omega)$ exhibit very strong and sharp spin-wave peaks.
Well above $T_{KT}$, $S^{xx}$ and $S^{zz}$
apparently display a central peak, and spin-wave signatures are still 
seen in $S^{zz}$. In addition, we also observed an almost dispersionless 
domain-wall peak at high $\omega$ below $T_{c}$(Ising transition), 
where long-range order 
appears in the staggered chirality. Above $T_{c}$, the domain-wall
peak disappears for all $q$. The lineshape of these peaks is captured 
reasonably well by a Lorentzian form. 
Using a dynamic finite-size scaling theory, 
we determined the dynamic critical exponent $z$ = 1.002(3). We found
that our results demonstrate the consistency of the dynamic finite-size 
scaling theory for the characteristic frequeny $\omega_{m}$ and 
the dynamic structure factor $S({\bf q},\omega)$ itself.
\end{abstract}
                  
\maketitle

\section{Introduction}
\label{sec0}

The study of the dynamic properties of classical spin systems has been
extensively carried out using both theoretical and simulational approaches.
In their work on the theory of dynamic critical phenomena\cite{H-H}, 
Hohenberg and Halperin proposed a number of different dynamic universality 
classes based upon the conservation laws.
The dynamic critical behavior is describable in terms of a dynamic 
critical exponent $z$, which gives rise to different dynamic universality 
classes: $z$ depends on the 
conservation laws, lattice dimension, and the static critical exponents.
Now spin dynamics simulations have become a mature method for probing 
the time dependent behavior of magnetic systems (see Ref. 2 for a recent 
review). This approach has resulted 
in high-quality dynamic critical exponent estimates for numerous models. 
In a recent high-resolution spin-dynamics study,\cite{Tsai} 
Tsai {\it et al.} made direct, quantitative comparison of 
both the dispersion curve and the lineshapes obtained from their simulation
data with recent experimental results for RbMnF$_{3}$ using the Heisenberg 
antiferromagnet model and a newly developed approach 
of higher order decomposition time integration methods\cite{KBL} and obtained 
a good agreement.

In this paper we consider the classical two-dimensional antiferromagnetic 
$XY$ model on a triangular lattice (TAFXY). 
This frustrated 2$D$ spin system has received 
much attention during the last decade\cite{Lee,Benakli,Miya,DHLee,SLee,Luca,Step}. The TAFXY model displays rich low-temperature phase structures 
and critical phenomena because frustration introduces additional discrete
symmetries resulting from the chiral degrees of freedom. The ground states
of this model are composed of three interpenetrating 
sublattices with lattice vectors of 
length $\sqrt{3}$. Spins on each sublattice are ferromagnetically ordered
and spins on different sublattices are oriented $\pm 2 \pi/3$
with respect to each other. The TAFXY has a continuous U(1) symmetry associated
with global spin rotations and a discrete $Z_{2}$ symmetry because of the 
double degeneracy of the ground-state chirality configurations.
This model has two
order parameters: the staggered in-plane magnetization and the staggered 
chirality. There have been ongoing controversies concerning phase transitions
and the nature of the phase transitions. Some\cite{Lee,Benakli} suggested that the system 
undergoes a single phase transition, generally yielding non-Ising critical 
exponent; however, others\cite{Miya,DHLee,SLee,Luca,Step} supported 
a double phase transition scenario 
with a Kosterlitz-Thouless(KT)-like transition followed by a second-order 
chirality-lattice melting transition at a slightly higher temperature.
In addition to spin waves and vortices, which are fundamental excitations 
in the ferromagnetic or antiferromagnetic $XY$ model on a bipartite lattice, 
the triangular antiferromagnetic $XY$ model has another
excitations associated with domain-wall formation between two different 
ground states in view of the chirality configuration of 
ground states\cite{DHLee}.

A good experimental realization of the frustrated antiferromagnetic 
$XY$ model on a stacked triangular lattice is ABX$_{3}$ type 
halides (like CsNiCl$_{3}$, CsMnBr$_{3}$, CsCuCl$_{3}$)\cite{HSG,VPP}. 
The magnetic B$^{2+}$ ions form a triangular antiferromagnet within 
the $ab$ planes and a ferromagnetic coupling along the $c$ direction 
produces three dimensional order.

While the static critical behavior of the TAFXY model has been the subject 
of much interest during the last decade, the theoretical study of 
dynamic critical behavior has not been carried out 
for the TAFXY model. 

In the present work, we have studied the dynamic 
behavior of the TAFXY model using Monte Carlo and spin-dynamic methods(MC-SD).
Although in the usual planar $XY$ model each spin has two components, here we use the 
three-component $XY$ model in order to study the real dynamics. These 
two models belong to the same static universality class\cite{Hans,Nho},
but the planar $XY$ model will not have true dynamical behavior.\cite{H-H}

The estimates for $T_{KT}$ and $T_{c}$ used here are taken from a recent 
high-precision MC study where
the critical temperature $T_{c}$ associated with the chirality phase 
transition and the KT transition temperature $T_{KT}$ associated with 
unbinding of vortex pairs have been determined as 
$T_{c}$ = 0.412(5)$J/k_{B}$ and $T_{KT}$ = 0.403(1)$J/k_{B}$.\cite{Luca}

\section{Model and simulation Method}
\label{sec1}

We will first briefly describe the model and the numerical method used
here and show how the dynamic structure factor is computed.
The classical antiferromagnetic $XY$ model is described by
the following form:

\begin{equation}
H = J\sum_{<ij>}(S^{x}_{i}S^{x}_{j}+S^{y}_{i}S^{y}_{j}), 
\end{equation}

\noindent
where the summation is over all nearest neighbors, ${\bf S}_{i}$ = 
$(S^{x}_{i},S^{y}_{i},S^{z}_{i})$ is a three-dimensional classical 
spin of unit length at site $i$, and $J $ is the positive (antiferromagnetic) 
coupling constant between nearest-neighbor pairs of spins. 
We consider triangular lattices of size $L \times L$ 
along the primitive vector directions($\hat{e}_{x},\frac{1}{2}\hat{e}_{x}+\frac{\sqrt{3}}{2}\hat{e}_{y})$, containing $N$ = $L^{2}$ sites 
and $2N$ elementary triangles. In our calculations, 
periodic boundary conditions are applied along the primitive vector 
directions.

 We use a hybrid Monte Carlo procedure which consists of a combination of 
the Metropolis update\cite{Metro} and 
the over-relaxation algorithm\cite{FRB} ( see also Ref. 18).
One hybrid Monte Carlo step consists of two Metropolis and eight 
over-relaxation updates\cite{Chen,Bunker}. Using this hybrid algorithm, 
we generated approximately 700-8500 equilibrium configurations
at a given temperature. Typically 1000 hybrid Monte Carlo steps were 
used to generate each equilibrium configuration, which is then
evolved using the equations of motion given by:

\begin{equation}
\frac{d}{dt}{\bf S}_{i} = \frac{\partial H}{\partial {\bf S}_{i}} \times 
{\bf S}_{i}.
\end{equation}

\noindent
Starting from a particular initial spin configuration, we performed 
numerical integration of these equations of motion using a recently 
developed 4th-order Suzuki-Trotter decomposition method\cite{KBL}, 
and the integration is carried out
to a maximum time $t_{max}$, typically of the order of $t_{max}J$ = 680
and $t_{cutoff}J$ = 600,
with a time step of $\delta t$ = 0.2$J^{-1}$. We compute the thermal average
of a time-dependent observable by averaging over all the values
of the observable obtained by evolving all the independent initial 
equilibrium configurations.

The dynamic structure factor $S^{kk}({\bf q},\omega)$ is the space-time 
Fourier transform of the position and time displaced spin-spin correlation 
function. It is defined for momentum transfer ${\bf q}$ and frequency transfer
$\omega$  as follows\cite{Chen}:

\begin{equation}
S^{kk}({\bf q},\omega) = \sum_{{\bf r},{\bf r'}}e^{i{\bf q}\cdot
({\bf r}-{\bf r'})} \int^{+\infty}_{-\infty}e^{i\omega t}C^{kk}({\bf r}
-{\bf r'},t)\frac{dt}{2\pi},
\end{equation}
where $k$ = $x,y$, or $z$ is the spin component and the space-displaced, time-displaced spin-spin correlation function is
given by 

\begin{eqnarray}
C^{kk}({\bf r}-{\bf r'},t-t') & = &  \langle S^{k}_{{\bf r}}(t)S^{k}_{{\bf r'}}(t')\rangle \nonumber \\
                              &   &  -\langle S^{k}_{{\bf r}}(t)\rangle
\langle S^{k}_{{\bf r'}}(t')\rangle.
\end{eqnarray}

The dynamic critical exponent $z$ can be determined by using the dynamic
finite-size scaling theory developed by Chen and Landau\cite{Chen}:

\begin{equation}
\frac{\omega S^{kk}_{L}({\bf q},\omega)}{\chi^{kk}_{L}({\bf q})}
= G^{kk}(\omega L^{z},qL),
\label{eq3}
\end{equation}
where we do not introduce a frequency resolution function in order to 
smoothen the effects of finite length of time integration because 
of very long integration times used in our spin dynamics.
$\chi^{kk}_{L}({\bf q})$ is the total 
integrated intensity given by

\begin{equation}
\chi^{kk}_{L}({\bf q}) = \int^{\infty}_{-\infty}S^{kk}_{L}({\bf q},\omega) d\omega.
\end{equation}

\noindent
The characteristic frequency $\omega^{kk}_{m}(qL)$ is a median frequency
determined by the constraint:

\begin{equation}
\frac{1}{2}\chi^{kk}_{L}({\bf q}) = \int^{\omega^{kk}_{m}}_{-\omega^{kk}_{m}}
S^{kk}_{L}({\bf q},\omega) d\omega.
\end{equation}

In the dynamic scaling theory, the finite-size scaling expression for the
median frequency $\omega^{kk}_{m}$ is given by

\begin{equation}
\omega^{kk}_{m} = L^{-z}\Omega(qL).
\label{eq8}
\end{equation}

\noindent
Using Eq. (\ref{eq8}), we estimate the dynamic critical exponent $z$
from the slope of a graph of log($\omega^{kk}_{m}$) versus log($L$) at
fixed value of $qL$, and we test the dynamic finite-size scaling theory
using Eq. (\ref{eq3}).
 
\section{Results and Discussion}
\label{sec2}

We first compare the behavior of the dynamic structure factor $S^{kk}(q,\omega)$ at several different temperatures, where
$k$ refers to the $x$ (in-plane) or $z$ (out-of-plane) component and
we were limited to the [11] reciprocal lattice direction, i.e. ${\bf q}=(q,q)$
with $q$ determined by the periodic boundary conditions, 

\begin{equation}
q=\frac{4\pi}{L}n_{q},  \hspace{.3in} n_{q}=1,2,...,\frac{L}{3}.
\end{equation}

\begin{figure}
\begin{picture}(0,400)(0,0)
\put(-140,-20){\includegraphics{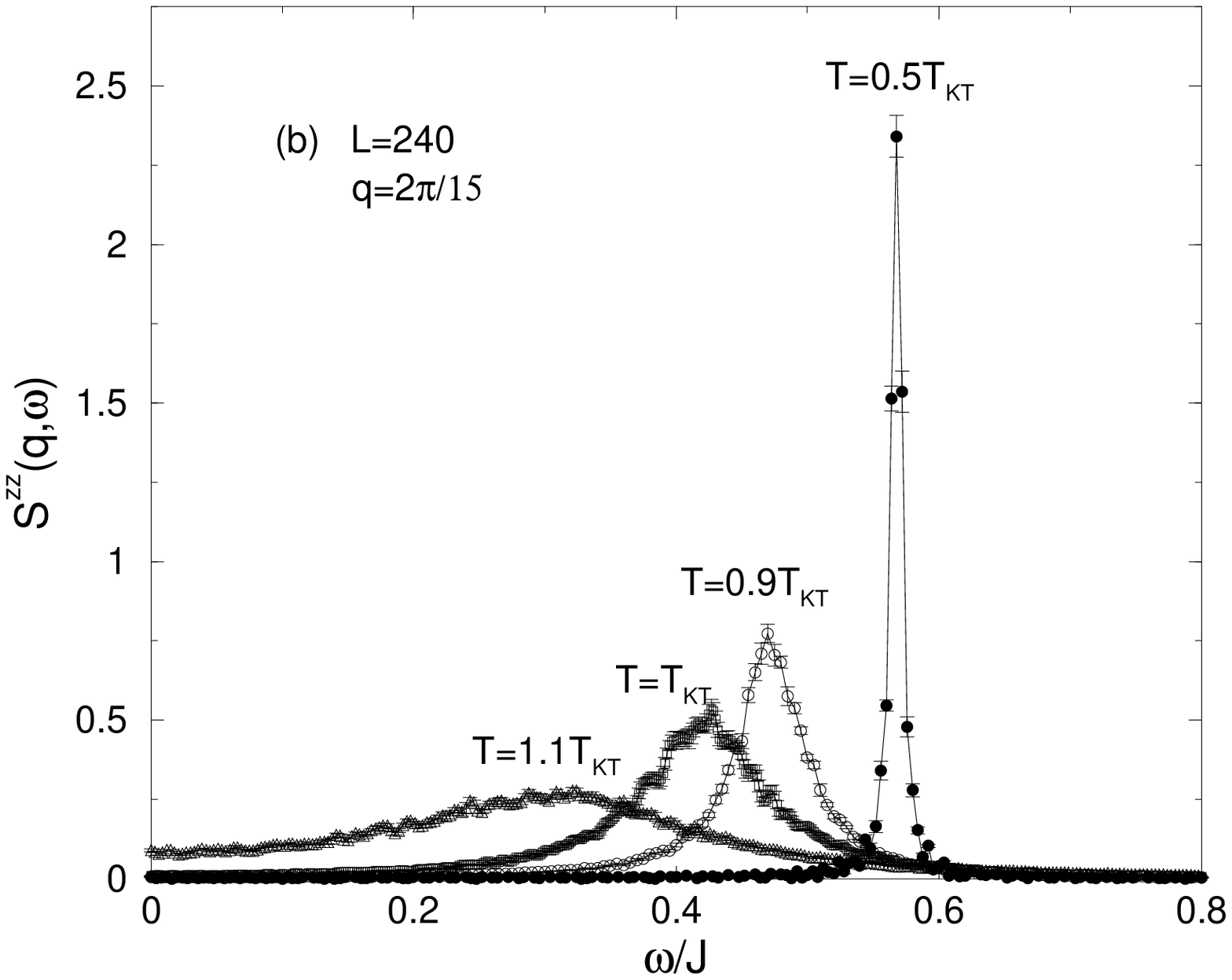}}
\put(-140,180){\includegraphics{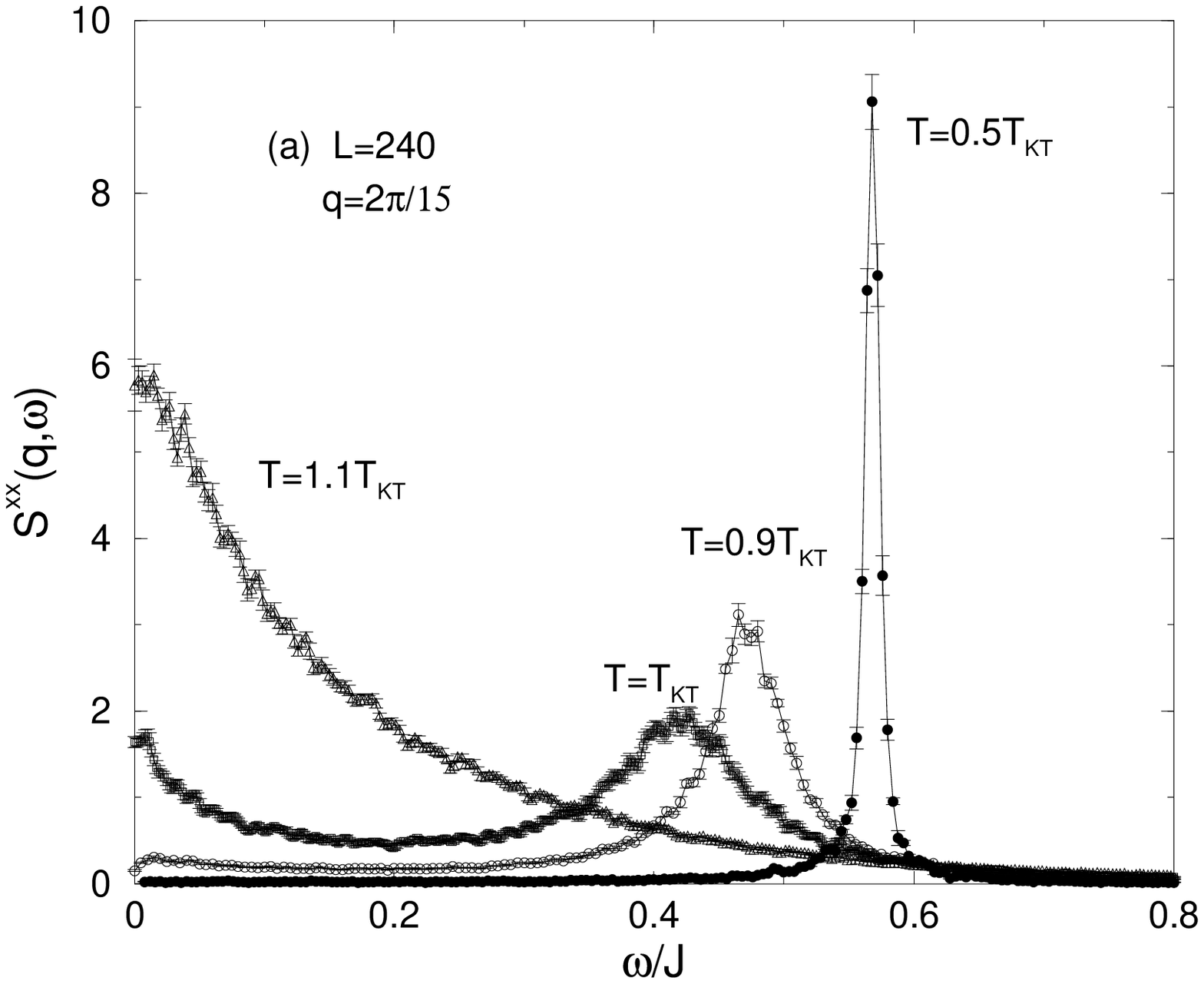}}
\end{picture}
\caption{Temperature dependence of the dynamic structure factor 
$S^{kk}(q,\omega)$. L=240 and $q$= 2$\pi$/15 in the [11] direction
in all cases: (a) in-plane component; (b) out-of-plane component. 
}
\label{fi-1}
\end{figure}

Figure 1 shows the temperature dependence of the dynamic 
structure factor $S^{kk}(q,\omega)$ for a lattice size of $L$=240 and 
a particular momentum $q$=$2\pi/15$. For $T \leq T_{KT}$ our results for the dynamic structure 
factor $S^{xx}$ show a very strong spin-wave and a central peak which are 
visible as pronounced peaks at the spin-wave frequency $\omega (q)$ and at
$\omega =0$, respectively. $S^{zz}$ has structures with 
less intensity than $S^{xx}$ (note the change in scale between 
Figs. 1(a) and 1(b)).
There is a sharp spin-wave peak and no central peaks are visible in $S^{zz}$.
The positions of the spin-wave peaks are the same for $S^{xx}$ and $S^{zz}$
and as the temperature increases,
the spin-wave peak broadens, its position moves towards 
lower $\omega$, and its amplitude decreases.
Above $T_{KT}$, $S^{xx}$ and $S^{zz}$ apparently display a central peak,
but spin-wave signatures are still visible in $S^{zz}$.
The same qualitative behavior was observed in spin-dynamics simulations of the $2D$ 
classical $XY$ model on a square lattice\cite{Hans}.

\begin{figure}
\begin{picture}(0,200)(0,0)
\put(-140,-20){\includegraphics{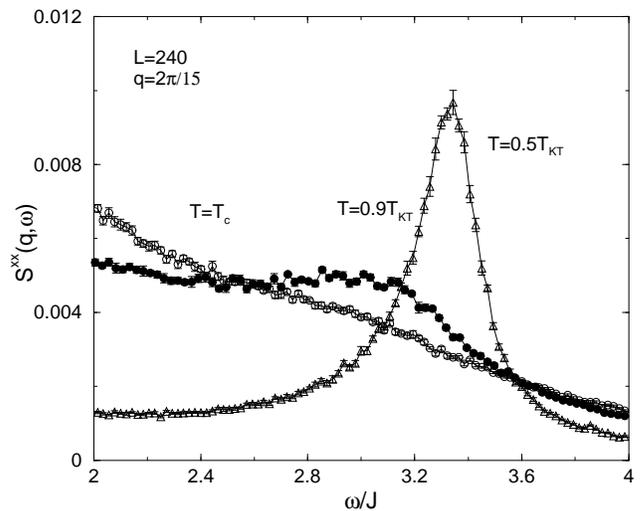}}
\end{picture}
\caption{Temperature dependence of the dynamic structure factor 
$S^{xx}(q,\omega)$ at high frequencies. 
L=240 and $q$= 2$\pi$/15 in the [11] direction in all cases.
For clarity, a frequency resolution function is used to smoothen
the oscillations due to the finite $t_{max}$. We can see a domain-wall
peak for $T < T_{c}$. The peak broadens as the temperature
increases. However, the domain-wall peak disappears at $T \geq T_{c}$.
}
\label{fi-2}
\end{figure}

\begin{figure}
\begin{picture}(0,200)(0,0)
\put(-140,-20){\includegraphics{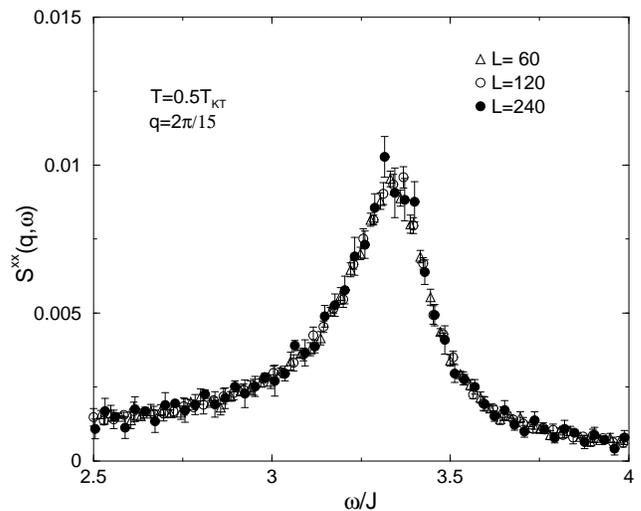}}
\end{picture}
\caption{Lattice size dependence of dynamic structure factor 
$S^{xx}(q,\omega)$ around the domain-wall peak position at fixed momentum 
$q$= 2$\pi$/15 in the [11] direction at $T = 0.5T_{KT}$. 
The intensity of the domain-wall peak 
does not depend on lattice size and the data essentially collapse 
onto a single curve.
}
\label{fi-3}
\end{figure}

As has been pointed out by Lee {\it et al.}\cite{DHLee}, in addition to spin waves,
there is another type of elementary excitation associated with domain-wall
formation between regions with opposite staggered chirality. In order to 
observe the dynamic effects of the destruction of chirality order, we investigate 
the temperature dependence of the dynamic structure factor $S^{xx}$ at high frequencies. We illustrate this in Fig. 2. The intensity of the domain-wall peak is $10^{-2}$ of that of the spin-wave peak at small $q$ ( note the change 
in scale between Figs. 1(a) and 2.). The relative intensity decreases at 
large $q$. For clarity, a frequency resolution function 
with $\delta\omega$= 1.2$\pi/t_{cutoff}$ is used to smoothen 
the oscillations due to the finite $t_{max}$.
Below $T_{c}$, where long-range order appears in the staggered chirality,
we observe an almost dispersionless domain-wall peak at high $\omega$ typical
of an Ising model.
As the temperature approaches the chirality transition $T_{c}$ from below,
the domain-wall peak broadens and its relative intensity decreases. 
The domain-wall peak disappears for all $q$ above $T_{c}$. This result
indicates the presence of an Ising-like phase transition connected with
the loss of chirality order.
Fig. 3 shows  $S^{xx}(q,\omega)$ for three different system sizes 
with the same value of $q = 2\pi/15$ at high $\omega$
in order to see the finite-size effects of the domain-wall peak.
The intensity of the domain-wall peak does not depend on lattice size, 
the data essentially collapse onto a single curve, and finite-size effects 
are not visible,
whereas the intensity of the spin-wave peak depends strongly on lattice size.

\begin{figure}
\begin{picture}(0,400)(0,0)
\put(-140,180){\includegraphics{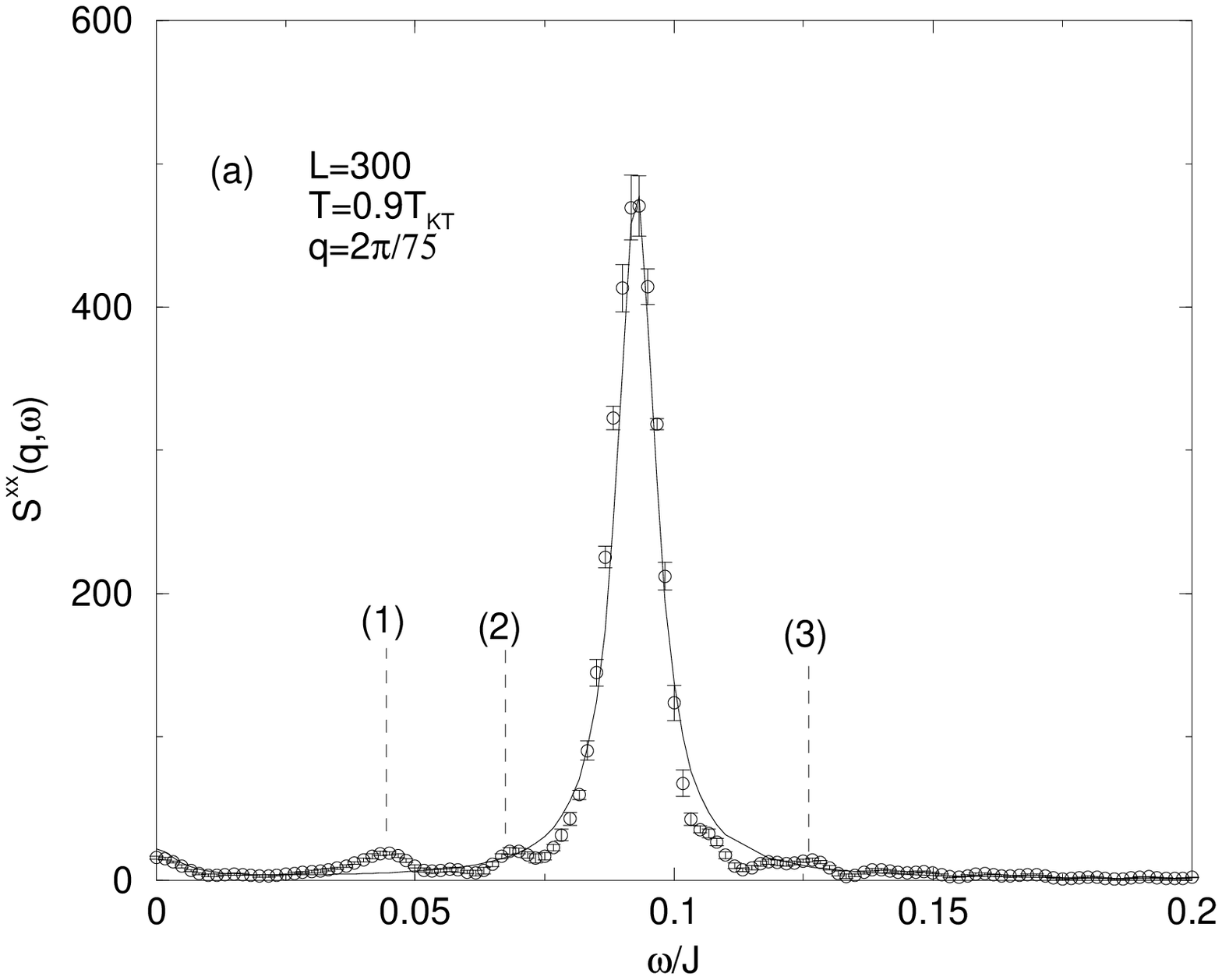}}
\put(-140,-20){\includegraphics{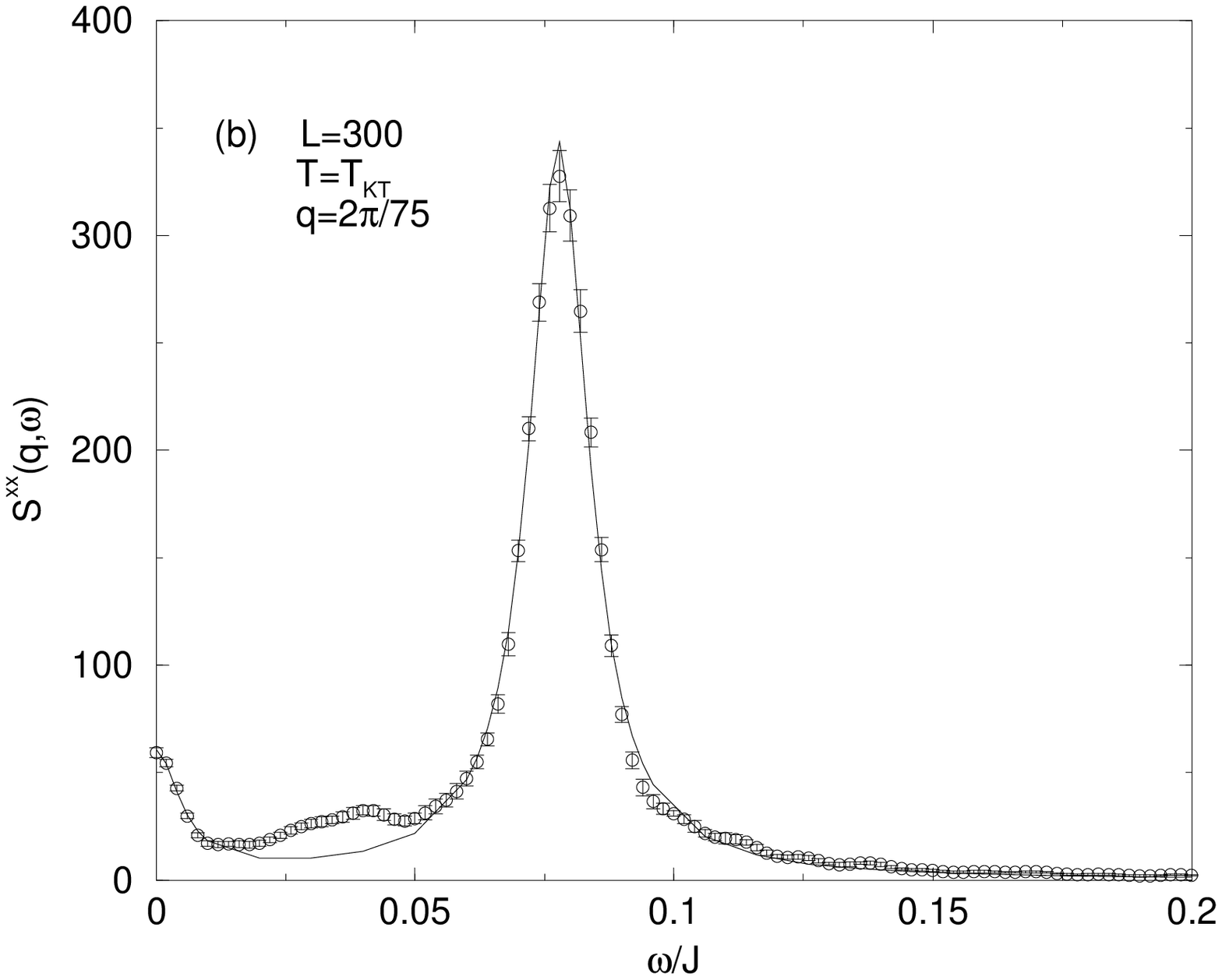}}
\end{picture}
\caption{In-plane component $S^{xx}$ of the dynamic structure factor
for $L=300$. (a) $T$= 0.9$T_{KT}$;
(b) $T$= $T_{KT}$ at $q$=2$\pi$/75 in the [11] direction in all cases. 
The symbols represent simulation data
and the solid line is a fit with the multiple Lorentzian function given in Eq. 
(\ref{eq-fit}).}
\label{fi-4}
\end{figure}

The in-plane dynamic structure factor $S^{xx}({\bf q},\omega)$ for
the 2$D$ $XY$ model below the topological transition temperature $T_{KT}$ 
was analyzed by Villain\cite{Vn}, Moussa and Villain\cite{MV}, and 
Nelson and Fisher\cite{NF}.
They found that $S^{xx}({\bf q},\omega)$ has spin-wave peaks diverging 
at the spin-wave frequency $\omega_{p}({\bf q})$.  
Menezes {\it et al.}\cite{Mz} calculated $S({\bf q},\omega)$ and found 
a diverging spin-wave peak
similar to that of Nelson and Fisher and a logarithmically diverging 
central peak. However Evertz and Landau\cite{Hans} found that their MC-SD data for
the shape of $S({\bf q},\omega)$ is not well described around 
the spin-wave peak by the above theoretical predictions. 

\begin{figure}
\begin{picture}(0,200)(0,0)
\put(-120,-10){\includegraphics{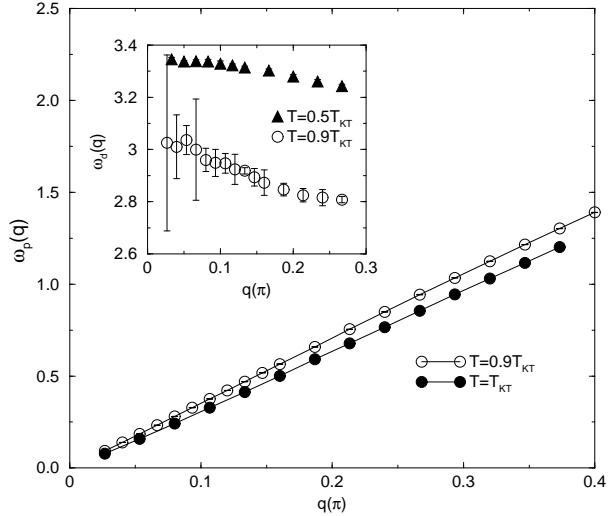}}
\end{picture}
\caption{Dispersion relation $\omega_{p}(q)$ of the spin-wave peak in
$S^{xx}(q,\omega)$ for the [11] direction as a function of momentum.
The inset shows the dispersion relation $\omega_{d}(q)$ of the domain-wall
peak.}
\label{fi-5}
\end{figure}

\begin{figure}
\begin{picture}(0,200)(0,0)
\put(-140,-20){\includegraphics{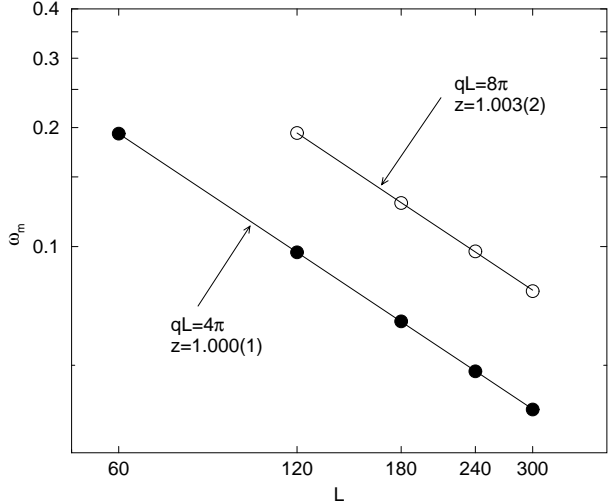}}
\end{picture}
\caption{Finite-size scaling plot for the median frequency $\omega^{xx}_{m}$
at $T = T_{KT}$.
The dynamic critical exponent value obtained from the slope
of the log-log plot is $z$=1.000(1) for $n$=1 using all lattice sizes
and $z$=1.003(2) for $n$=2 using the largest four lattice sizes.
The solid lines display a fit to a straight line.
Statistical errors of the data are smaller than the symbol sizes.}
\label{fi-6}
\end{figure}

Mertens {\it et al.}\cite{Ms1,Ms2} calculated $S({\bf q},\omega)$ above $T_{KT}$ and
found a Lorentzian central peak for $S^{xx}$ and a Gaussian central peak 
for $S^{zz}$. Very recently, Wysin {\it et al.}\cite{Wy} calculated $S({\bf q},\omega)$
 and found nondivergent spin-wave peaks and weak peaks on both sides of 
the spin-wave peaks. Below $T_{KT}$, these theoretical approaches motivated 
us to fit the lineshape of the dynamic structure factor $S^{xx}$ 
to a Lorentzian form. We found that the lineshape of $S^{xx}$ is 
reasonably well captured by a multiple Lorentzian form:

\begin{eqnarray}
S^{xx}(q,\omega)&=&\frac{A\Gamma^{2}_{1}}{\Gamma^{2}_{1}+\omega^{2}}+
      \frac{B\Gamma^{2}_{2}}{\Gamma^{2}_{2}+(\omega-\omega^{2}_{s})}+
      \frac{B\Gamma^{2}_{2}}{\Gamma^{2}_{2}+(\omega+\omega^{2}_{s})} \nonumber \\
                & & +\frac{C\Gamma^{2}_{3}}{\Gamma^{2}_{3}
      +(\omega-\omega^{2}_{d})}+
      \frac{C\Gamma^{2}_{3}}{\Gamma^{2}_{3}+(\omega+\omega^{2}_{d})},
\label{eq-fit}
\end{eqnarray}
where the first term corresponds to the central peak, the next two terms
are from the spin-wave creation and annihilation peaks 
at $\omega$=$\pm \omega_{s}$, and the last two terms are contributions from 
the domain-wall peaks at $\omega$=$\pm \omega_{d}$. In order to fit 
the simulated lineshapes to the Lorentzian form Eq. (\ref{eq-fit}), 
we used a frequency range $0 \leq \omega \leq 5 $. We find that the 
Lorentzian lineshapes fit our simulation data reasonably
well. Illustrations of the fits using Eq. (\ref{eq-fit}) to the simulated 
lineshapes at $T$= 0.9$T_{KT}$ and at $T_{KT}$ are shown in Fig. 4.
In the figure the symbols represent our simulation data and the solid line
is a fit with the Lorentzian function given in Eq. (\ref{eq-fit}).
In Fig. 4(a) and (b), we find additional small peaks on both sides 
of the spin-wave peak. One simple explanation which is consistent with the data
is the presence of two-spin-wave peaks. 
We could see that the position of these extra peaks
corresponded to frequencies of two spin-wave addition and difference peaks
(marked as (1), (2), and (3) in Fig. 4(a)).
These dynamical features have been noticed in earlier MC-SD simulations\cite{Tsai,Hans,MDPL} and
a theoretical approach\cite{Wy}.
As we did not generalize Eq. (\ref{eq-fit}) to include the extra peaks,
this extra structure causes the lineshape to depart from a Lorentzian form.

\begin{figure}
\begin{picture}(0,400)(0,0)
\put(-140,-20){\includegraphics{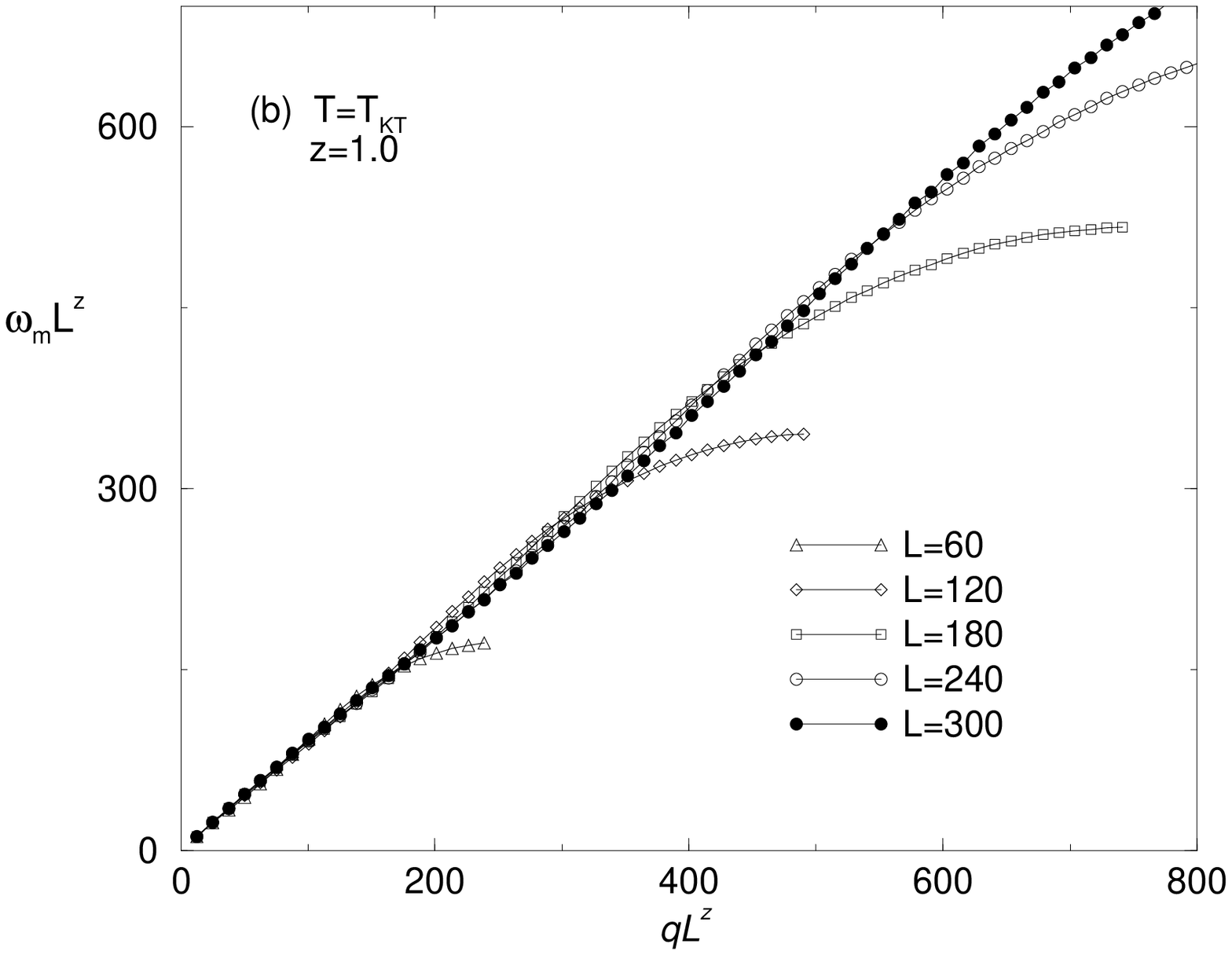}}
\put(-140,180){\includegraphics{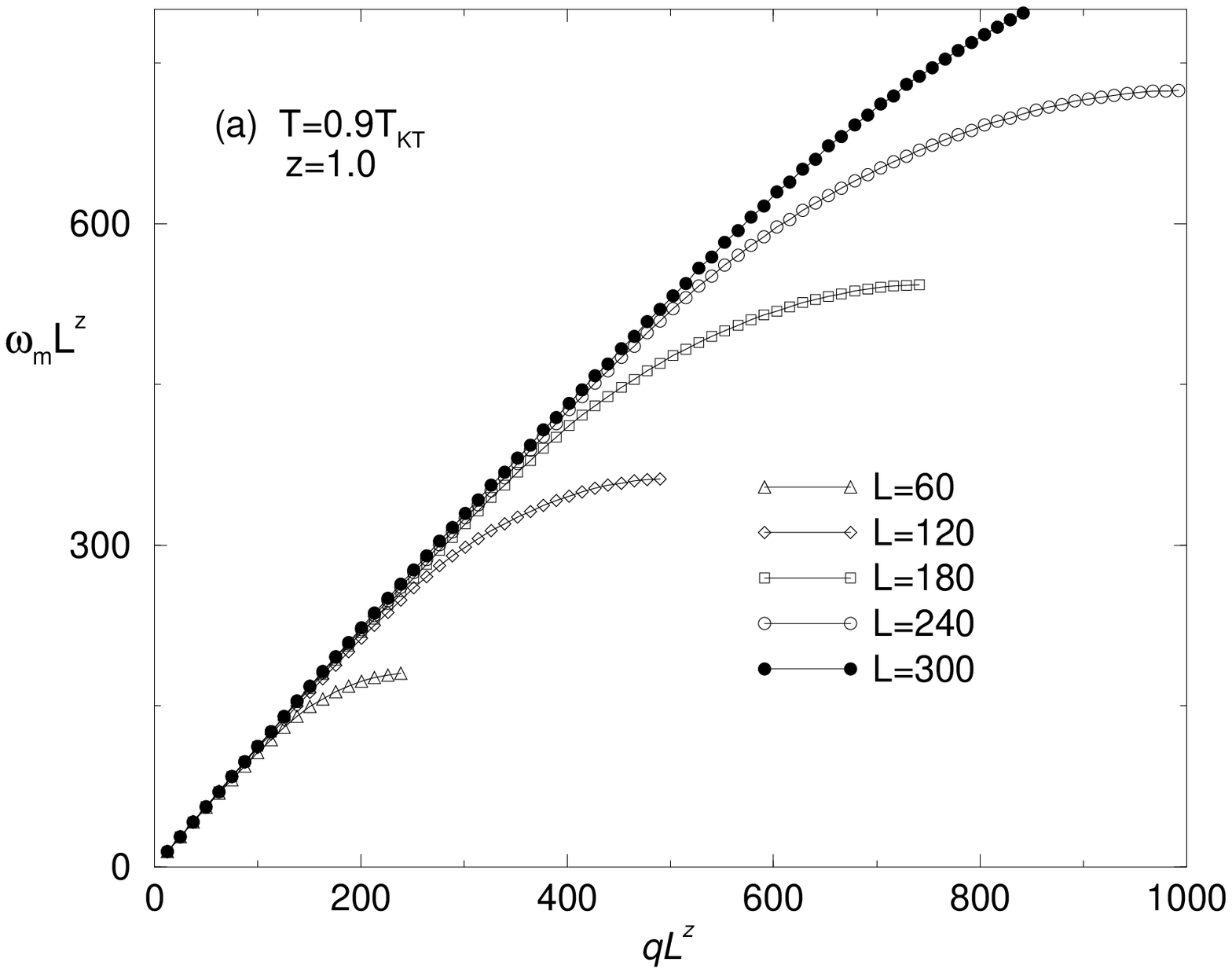}}
\end{picture}
\caption{Finite-size scaling of the median frequency( $\omega^{xx}_{m}L^{z}$
as a function of $qL^{z}$) at (a) $T$ = 0.9$T_{KT}$; (b) $T_{KT}$ 
using $z$=1.0.
The data for $\omega^{xx}_{m}$ show good scaling behavior for $T \leq T_{KT}$.
Statistical errors of the data are smaller than the symbol sizes.}
\label{fi-7}
\end{figure}

The dispersion relations $\omega_{p}(q)$ of the spin-wave peak and 
$\omega_{d}(q)$ of the domain-wall peak can be obtained from the above 
Lorentzian fit. Fig. 5 shows the dispersion curves as a function of momentum.
The inset shows the dispersion relation $\omega_{d}(q)$ at $T=0.5T_{KT}$
and $T=0.9T_{KT}$, 
where obtaining a good $\omega_{d}(q)$ is difficult because of the small peak 
height and the broadening of the domain-wall peak (see Fig. 2).
As expected, $\omega_{p}(q)$ is linear for small $q$ and $\omega_{d}(q)$
is approximately constant.

The dynamic critical exponent $z$ can be extracted from the finite-size 
scaling of the characteristic frequency $\omega_{m}$, 
as described in a previous section.
We calculated $\omega_{m}^{xx}(qL)$ using $S^{xx}$ with $qL$=4$\pi$($n$=1)
and $qL$=8$\pi$($n$=2) for several lattice sizes at $T_{KT}$.
The log-log plot of the characteristic frequency $\omega_{m}^{xx}$ as a 
function of the lattice size $L$ is shown in Fig. 6 where the estimated 
error bars for individual points are smaller than the symbol sizes. 
The solid lines display a fit to linear behaviors.
From the slope of a log($\omega_{m}^{xx}$) vs log($L$) graph, we obtained the dynamic critical exponent $z$
= 1.000$\pm$0.001 for $n$=1 using all lattice sizes and $z$= 1.003$\pm$0.002 
for $n$=2 using only the largest four lattice sizes. Within their respective 
error bars, the two estimates for $z$ agree. Our final estimate for the
dynamic critical exponent is $z$ = 1.002(3), where the error bar reflects 
the fluctuations in the different estimates of $z$.
For comparison, we proceeded to estimate the dynamic critical exponent $z$
at $T_{c}$. We obtained the dynamic critical exponent $z$=1.051$\pm$0.005 
for $n$=1 (not shown).
We then calculated $\omega^{kk}_{m}(qL)$ using $S^{kk}$
for all $q$ in the [11] direction and all lattice 
sizes $L$ and we graphed $\omega^{kk}_{m}L^{z}$ as a function of $qL^{z}$ 
for $T$ $\leq$ $T_{KT}$ using $z$=1.0. This graph is shown in Fig. 7. 
In the figure, statistical errors of the data are smaller than the symbol 
sizes. As we know, the dispersion curve flattens
for each finite lattice size when $q$ becomes large; therefore, the data
start to move away from the asymptotic behavior at progressively larger 
values of $qL$. The data for $\omega_{m}^{xx}$ show good scaling behavior 
for $T \leq T_{KT}$. The out-of-plane characteristic frequency 
$\omega_{m}^{zz}$ has the same scaling behavior as the in-plane component
at temperatures $T \leq T_{KT}$. Interestingly, at $T = T_{c}$,
$\omega_{m}^{zz}$ also shows the same good 
scaling behavior with $z$=1.0; however, we do not observe similar 
scaling behavior in 
$\omega_{m}^{xx}$ at $T = T_{c}$. The same behavior was observed 
by spin-dynamics simulation of the 2$D$ $XY$ model on a square lattice 
in a qualitative sense.\cite{Hans}

\begin{figure}
\begin{picture}(0,400)(0,0)
\put(-140,-20){\includegraphics{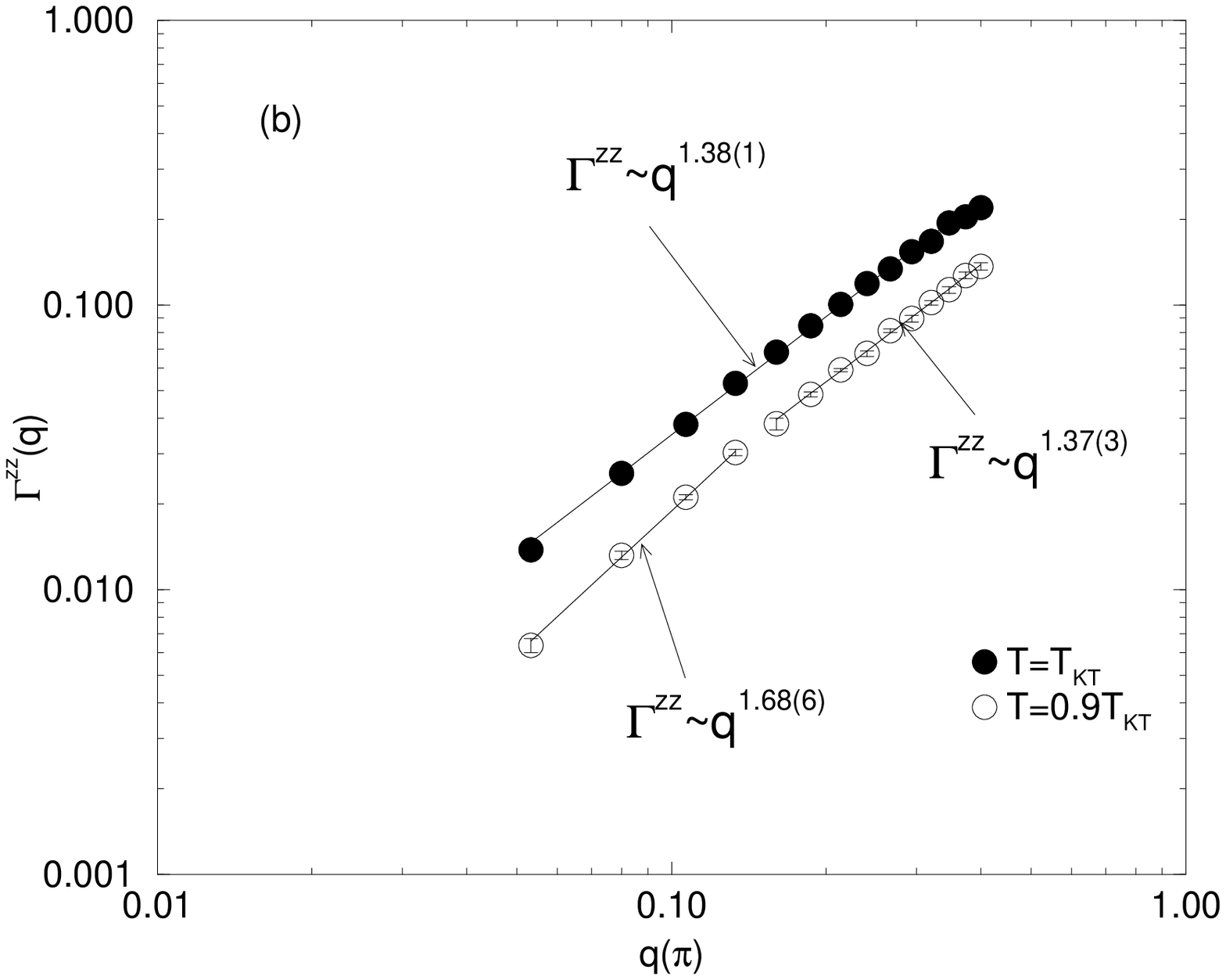}}
\put(-140,180){\includegraphics{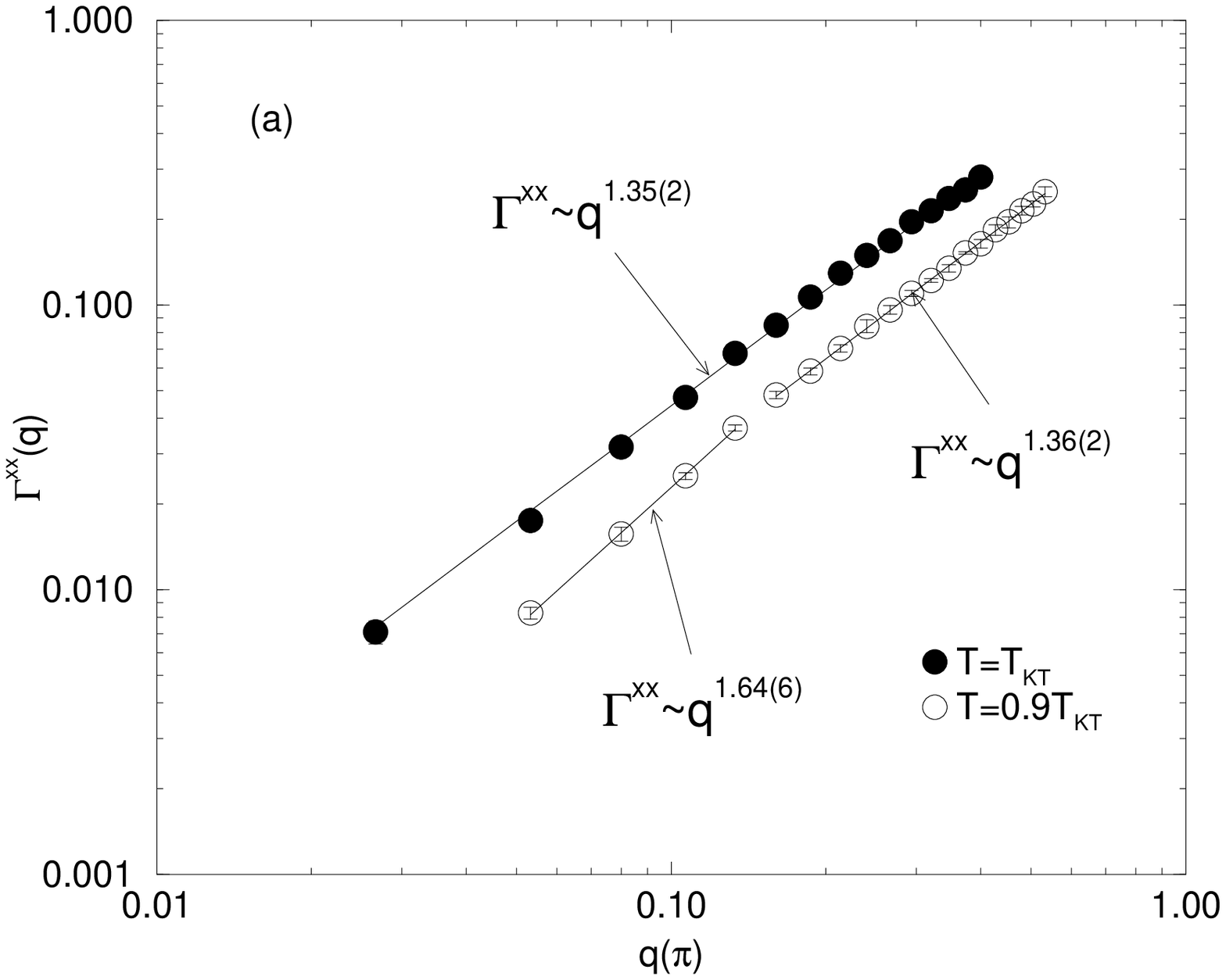}}
\end{picture}
\caption{Linewidths (a)$\Gamma^{xx}(q)$ and (b)$\Gamma^{zz}(q)$ of 
the spin-wave peak in $S^{kk}(q,\omega)$ for the [11] direction and $L=300$. 
The solid lines are linear fits to the data points.}

\label{fi-8}
\end{figure}

We had also attempted to estimate the $q$ dependence of the half-width 
$\Gamma^{kk}(q)$ of spin-wave peak in the critical region. 
Here we expect the simple form $\Gamma^{kk}(q) \sim q^{z}$. 
The log-log plot of the half-width $\Gamma^{kk}(q)$
from our simulations is shown in Fig. 8. For $\Gamma^{zz}(q)$, we estimated 
the error in the fitted parameters by fitting the lineshapes using three 
different ranges of frequency around the spin-wave peak. The fitted 
parameters varied when different frequency ranges were used in the fit.
For $\Gamma^{xx}(q)$, however, as there are three peaks( central, 
spin-wave, and domain-wall ) in $S^{xx}(q,\omega)$ below $T_{c}$, 
obtaining a good fitted parameter was more difficult than $\Gamma^{zz}(q)$.
First, the lineshape parameters are obtained from the fit to only 
one frequency range. Using these parameters, the central peak 
and the domain-wall peak are subtracted. 
The error is estimated by varying the size of the frequency 
window around the spin-wave peak.
In contrast to our expectation, we observed two different dynamic exponents
for two different regions of $q$ in $\Gamma^{kk}(q)$ at $T=0.9T_{KT}$.
In order to check our result, we calculated $\Gamma^{zz}(q)$ for the 
two-dimensional $XY$ model on a square lattice and obtained two different 
exponents as well(not shown).

In order to test the dynamic finite-size scaling theory of the dynamic
structure factor $S^{kk}$ itself through the use of Eq. (\ref{eq3}), we plotted 
$S_{L}^{kk}(q,\omega)$/$L^{z}\chi_{L}^{kk}(q)$ vs $\omega L^{z}$ for 
four lattice sizes with the dynamic critical exponent set to $z$=1.0.
For $qL$=4$\pi$ ($n$=1) and $T$=$T_{KT}$, the resulting scaling plot 
for $S^{xx}(q,\omega)$ is shown in Fig. 9. Data points collapse onto 
the same curve for sufficiently large $L$, namely lattice sizes $L\geq 240$.
Only the data from small lattices deviate systematically.
Scaling works quite well for all frequencies for lattice sizes $L$=240 and 300.
The corresponding plot for $S^{xx}(q,\omega)$ and $qL$=4$\pi$ 
at $T$=$T_{c}$ is displayed in Fig. 10. The dynamic structure factors for 
the largest three lattice sizes  fall onto a single curve within the 
bounds of their error bars. Our MC-SD data clearly show that 
even above $T_{KT}$ the dynamic structure factor scales quite well, 
consistent with the results of the 2$D$ $XY$ model on a square lattice 
by Evertz {\it et al.}\cite{Hans}. 
For the out-of-plane component $S^{zz}$, we do observe scaling (not shown).
Thus we can validate the dynamic finite-size scaling theory.

\begin{figure}
\begin{picture}(0,200)(0,0)
\put(-140,-20){\includegraphics{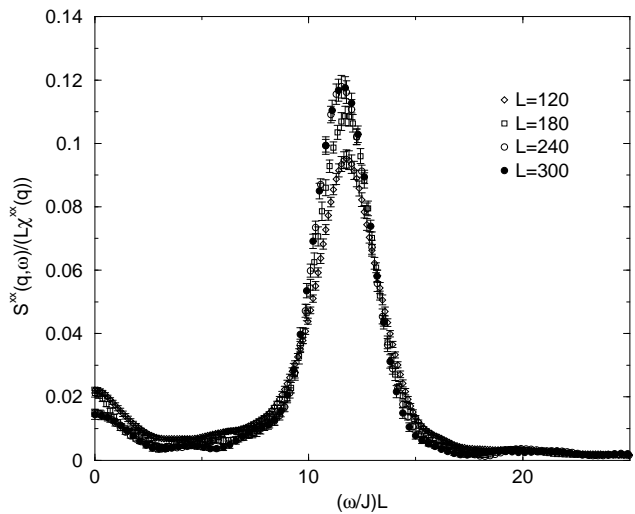}}
\end{picture}
\caption{Finite-size scaling of the dynamic structure factor $S^{xx}$ 
for $z$=1.0, $qL$= 4 $\pi$ in the [11] direction, and $T$=$T_{KT}$. 
Data points collapse onto the same curve for 
sufficiently large $L$. Only the data from small lattices deviate 
systematically.
}
\label{fi-9}
\end{figure}

\begin{figure}
\begin{picture}(0,200)(0,0)
\put(-140,-20){\includegraphics{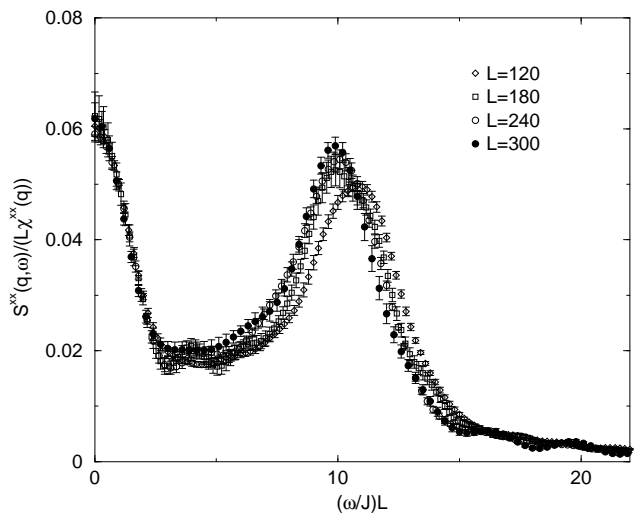}}
\end{picture}
\caption{Finite-size scaling of the dynamic structure factor $S^{xx}$
for $z$=1.0, $qL$= 4 $\pi$ in the [11] direction, and $T$=$T_{c}$. 
Data points collapse onto the same curve for 
sufficiently large $L$. Only the data from small lattices deviate 
systematically.
}
\label{fi-10}
\end{figure}

In summary, we have studied the dynamic critical behavior of the classical 
2$D$ antiferromagnetic planar($XY$) model on a triangular lattice using 
a combination of Monte Carlo methods and spin-dynamics techniques which 
use recently developed decomposition methods in order to increase the 
efficiency of the simulations.
We have calculated the dynamic structure factor $S^{kk}(q,\omega)$ at
temperatures below, at, and above $T_{c}$ with lattice sizes $L \leq 300$.

At $T \leq T_{KT}$, very strong, sharp spin-wave peaks occur
in both the in-plane and out-of-plane components of the dynamic structure factor $S(q,\omega)$.
As $T$ increases, the spin-wave peak widens and its position moves towards 
lower $\omega$. In addition to the spin-wave peak, $S^{xx}(q,\omega)$
displays a central peak and additional two-spin-wave peaks. No central 
peaks are visible in $S^{zz}(q,\omega)$. Above $T_{KT}$, $S^{xx}(q,\omega)$ 
and $S^{zz}(q,\omega)$
apparently display a central peak and spin-wave signatures are still 
in $S^{zz}$. In a qualitative sense, the same behavior was observed in 
spin-dynamics simulations of the 2$D$ $XY$ model 
on a square lattice\cite{Hans}.

Below $T_{c}$, where long-range order appears in the staggered chirality,
we observed an almost dispersionless domain-wall peak at high $\omega$.
The domain-wall peak broadens and its intensity decreases as the
temperature increases. However the intensity of the domain-wall peak does not 
depend on lattice size. The domain-wall peak disappears for all $q$ 
above $T_{c}$.

The lineshape of the central, spin-wave, and domain-wall peaks is captured 
reasonably well by a Lorentzian 
form. We estimated the dynamic critical exponent $z$ from the finite-size
scaling of the characteristic frequency $\omega_{m}$. Our determined value
is $z$=1.002(3) that is in agreement 
with that of the 2$D$ $XY$ model 
on a square lattice $z$=1.00(4)\cite{Hans}.

Finally, we have examined the dynamic finite-size scaling theory and 
we found that the characteristic frequency $\omega_{m}^{kk}(qL)$ and 
the dynamic structure factor $S^{kk}(q,\omega)$ itself scale very well.

\begin{acknowledgments}

We are indebted to Shan-Ho Tsai and H. K. Lee 
for helpful discussions.
This work was partially supported by NSF grant No. DMR-0094422.
Our simulations were carried out on IBM SP (Blue Horizon) at the San Diego 
Supercomputing Center and IBM SP at the University of Georgia.

\end{acknowledgments}


\begin{thebibliography}{lbl}
\bibitem[1]{H-H}
P. C. Hohenberg and B. I. Halperin, Rev. Mod. Phys. {\bf 49}, 435 (1977).
\bibitem[2]{L-K}
D. P. Landau and M. Krech, J. Phys.: Condens. Matter {\bf 11}, R179 (1999).
\bibitem[3]{Tsai}
S. H. Tsai, A. Bunker, and D. P. Landau, Phys. Rev. B {\bf 61}, 333 (2000).\bibitem[4]{KBL} 
M. Krech, A. Bunker, and D. P. Landau, Comp. Phys. Comm. {\bf 111}, 1 (1998);
J. Frank, W. Huang, and B. Leimkuhler, J. Comp. Phys. {\bf 133}, 160 (1997).
\bibitem[5]{Lee}
J. Lee, J. M. Kosterlitz, and E. Granato, Phys. Rev. B {\bf 43}, 11531 (1991).
\bibitem[6]{Benakli}
M. Benakli, H. Zheng, and M. Gabay, Phys. Rev. B {\bf 55}, 278 (1997).
\bibitem[7]{Miya}
S. Miyashita and H. Shiba, J. Phys. Soc. Jpn. {\bf 53}, 1145 (1984).
\bibitem[8]{DHLee}
D. H. Lee, J. D. Joannopoulos, J. W. Negele, and D. P. Landau, Phys. Rev. Lett.
{\bf 52}, 433 (1984); Phys. Rev. B {\bf 33}, 450 (1986).
\bibitem[9]{SLee}
S. Lee and K. C. Lee, Phys. Rev. B{\bf 57}, 8472 (1998).
\bibitem[10]{Luca}
L. Capriotti, R. Vaia, A. Cuccoli, and V. Tognetti,
Phys. Rev. B {\bf 58}, 273 (1998).
\bibitem[11]{Step}
W. Stephan and B. W. Southern, Phys. Rev. B {\bf 61}, 11514 (2000).
\bibitem[12]{HSG}
H. Serrano-Gonzalez, S. T. Bramwell, K. D. M. Harris, B. M. Kariuki, 
L. Nixon, I. P. Parkin, and C. Ritter, Phys. Rev. B {\bf 59}, 14451 (1999).
\bibitem[13]{VPP}
V. P. Plakhty, S. V. Maleyev, J. Kulda, J. Wosnitza, D. Visser, 
and E. Moskvin, Europhys. Lett, {\bf 48}, 215 (1999).
\bibitem[14]{Hans}
H. G. Evertz and D. P. Landau, Phys. Rev. B {\bf 54}, 12302 (1996).
\bibitem[15]{Nho}
K. Nho and E. Manousakis, Phys. Rev. B {\bf 59}, 11575 (1999).
\bibitem[16]{Metro}
N. Metropolis, A. W. Rosenbluth, M. N. Rosenbluth, A. M. Teller, and E. Teller,
J. Chem. Phys. {\bf 21} 1087 (1953).
\bibitem[17]{FRB}
F. R. Brown and T. J. Woch, Phys. Rev. Lett. {\bf 58}, 2394 (1987);\\
M. Creutz, Phys. Rev. D {\bf 36}, 515 (1987).
\bibitem[18]{BoL}
D. P. Landau and K. Binder, {\it Monte Carlo Simulations in Statistical Physics}(Cambridge University Press, Cambridge, 2000).
\bibitem[19]{Chen}
K. Chen and D. P. Landau, Phys. Rev. B {\bf 49}, 3266 (1994).
\bibitem[20]{Bunker}
A. Bunker, K. Chen, and D. P. Landau, Phys. Rev. B {\bf 54}, 9259 (1996).
\bibitem[21]{Vn}
J. Villain, J. Phys. (Paris) {\bf 35}, 27 (1974).
\bibitem[22]{MV}
F. Moussa and J. Villain, J. Phys. C {\bf 9}, 4433 (1976).
\bibitem[23]{NF}
D. R. Nelson and D. S. Fisher, Phys. Rev. B {\bf 16}, 4945 (1977).
\bibitem[24]{Mz}
S. L. Menezes, A. S. T. Pires, and M. E. Gouv$\hat{e}$a, Phys. Rev. B {\bf 47}, 12280 (1993).
\bibitem[25]{Ms1}
F. G. Mertens, A. R. Bishop, G. M. Wysin, and C. Kawabata, Phys. Rev. Lett. {\bf 59}, 117 (1987);Phys. Rev. B {\bf 39}, 591 (1989).
\bibitem[26]{Ms2}
F. G. Mertens, A. R. Bishop, M. E. Gouv$\hat{e}$a, and G. M. Wysin, J. de Physique C 
{\bf 8}, 1385 (1988); M. E. Gouv$\hat{e}$a, G. M. Wysin, A. R. Bishop, and F. G. 
Mertens, Phys. Rev. B {\bf 39}, 11840 (1989).
\bibitem[27]{Wy}
G. M. Wysin, M. E. Gouv$\hat{e}$a, and A. S. T. Pires, Phys. Rev. B {\bf 62}, 11585 (2000).
\bibitem[28]{MDPL}
M. Krech and D. P. Landau, Phys. Rev B {\bf 60}, 3375 (1999).
\end{thebibliography}
\end{document}